\documentclass [a4paper,fleqn, 12pt]{article}
\usepackage{graphicx}
\usepackage[small]{subfigure,epsfig}

\usepackage {amsmath} \usepackage{amssymb} \usepackage{cite}

\newcommand{\cn}

\begin{document}


\title
{Meromorphic exact solutions of the generalized Bretherton equation}

\author
{Nikolay A. Kudryashov, \and Dmitry I. Sinelshchikov, \and Maria V. Demina}

\date{Department of Applied Mathematics, National Research Nuclear University
MEPHI, 31 Kashirskoe Shosse,
115409 Moscow, Russian Federation}




\maketitle

\begin{abstract}

The generalized Bretherton equation is studied. The classification of the meromorphic traveling wave solutions for this equation is presented. All possible exact solutions of the generalized Brethenton equation are given.

\end{abstract}






\section{Introduction}

In recent papers \cite{Kudryashov2010, Kudryashov2010a}, we proved two theorems for possible presentation of solutions for nonlinear ordinary differential equations on the complex plane. These theorems allow us to have a classification of exact solutions for nonlinear differential equations and lead to a new method for constructing meromorphic exact solutions of some nonlinear ordinary differential equations. In the case one branch for expansion of the general solution in the Laurent series  we obtain the known methods for finding exact solutions which were developed in the last years \cite{Kudr88, Kudr90, Kudr90a, Kudr91, Kudr92, Parkes01, Parkes02, Kudr05, Biswas01, Vernov02, Kudr2010b, Vitanov01}. However for two or more branches of expansions for the general solution in the Laurent series we can look for new exact expressions for exact solutions of nonlinear differential equations.

The essence of approach from papers \cite{Kudryashov2010, Kudryashov2010a} consists in looking exact solutions of nonlinear ordinary differential equations by comparing the Laurent series for the general solution of ordinary differential equation and the Laurent series for functions that can be a solution to the original equation. Taking this approach into account the meromorphic exact solution of third order differential equation were found in \cite{Kudryashov2010}. We also apply our method for constructing the exact solutions of the Kawahara equation expressed via the Weierstrass elliptic function in \cite{Kudryashov2010a}.

In this paper we look for all possible  traveling wave solutions of nonlinear evolution equation which is the Bretherton equation and takes the form

\begin{equation}
u_{tt}+\alpha\,u_{xx}+\beta\,u_{xxxx}+\delta\,u^{m}+\gamma\,u^{n}=0,\quad n>m, \quad n\neq1
\label{Bretherton}
\end{equation}


Using traveling wave $u(x,t)=y(z)$, where $z=x-C_0\,t$ in \eqref{Bretherton} we have the nonlinear ordinary differential equation
\begin{equation}
(\alpha-C_{0}^{2})\,y_{zz}+\beta\,y_{zzzz}+\delta\,y^{m}+\gamma\,y^{n}=0
\label{rrBretherton}
\end{equation}

Integrating equation \eqref{rrBretherton} once with respect to $z$ we obtain
\begin{equation}
\frac{\alpha-C_{0}^{2}}{2}\,y_{z}^{2}+\beta(y_{z}\,y_{zzz}-\frac{1}{2}\,y_{zz}^{2})+
\frac{\delta}{m+1}\,y^{m+1}+\frac{\gamma}{n+1}\,y^{n+1}+C_{2}=0,
\label{rBretherton}
\end{equation}
where $C_{2}$ is constant of integration.

In work \cite{Kudryashov2010a} the full classes of meromorphic traveling wave solutions of \eqref{rrBretherton} were have obtained at $m=1$ in the cases $n=2$, $n=3$ and $n=5$.

The aim of this work is to search for meromorphic the traveling wave solutions of Eq.\eqref{Bretherton} at any $n$ and $m=1$. We show that the elliptic solutions exist at $n=2$, $n=3$, $n=5$ and at $n=\frac73$. We also study equation \eqref{Bretherton} in the case  $n=5$ at $m=2,3,4$. To achieve our goal we will use approach suggested in \cite{Kudryashov2010,Kudryashov2010a}.

The outline of this paper is as follows. In Section 2 we show that Eq.\eqref{Bretherton} has elliptic traveling wave solutions at $m=1$ if and only if $n=2$, $n=3$, $n=5$. In Sections 3 and 4 we obtain meromorphic traveling wave solutions of Eq.\eqref{Bretherton} at $m\neq1$.

\section{Exact solutions of Eq. \eqref{rBretherton} at any $n$ in the case $m=1$.}

Consider the generalized Bretherton equation in form \eqref{rrBretherton}. Denoting $\alpha-C_{0}^{2}=\sigma$ we have
\begin{equation}
\sigma\,y_{zz}+\beta\,y_{zzzz}+\delta\,y+\gamma\,y^{n}=0
\label{rrBretherton_1}
\end{equation}
Without loss of the generality we take
\begin{equation}
\beta=-1,\quad \gamma=1
\end{equation}
From \eqref{rrBretherton_1} we have
\begin{equation}
\sigma\,y_{zz}-\,y_{zzzz}+\delta\,y+y^{n}=0
\label{rrBretherton_2}
\end{equation}

Let us prove the following theorem.

{\textbf{Theorem}{In the case $\sigma \neq 0$ and $\delta \neq 0$ there are elliptic solutions of equation \eqref{rrBretherton_2} if and only if $n=2, 3, 5$.}}

To prove this theorem we use the approach by paper \cite{Kudryashov2009}, which allows us to remove arbitrary exponent $n$ from Eq.\eqref{rrBretherton_2}.
Let us introduce the B\"{a}cklund transformations between solutions of Eq. \eqref{rrBretherton_2} and transformed equation which has solution with pole of the first order.

Taking into account the transformations
\begin{equation}
\begin{gathered}
w=\frac{y_{z}}{y},\\
y^{n-1}=-\delta+(w_{zzz}+3\,w_{z}^{2}+4\,w\,w_{zz}+6\,w^{2}\,w_{z}+w^{4})-\beta(w_{z}+w^{2})
\label{Bck}
\end{gathered}
\end{equation}
from Eq. \eqref{rrBretherton_2} we have
\begin{equation}
\begin{gathered}
3\,w  \left(n-5 \right) w_{z}^{2}+ \left( 3\,\sigma\,w+6\,n w ^{3}-10\,w ^{3}-10\,w_{zz}-n \sigma  w \right) w_{z}+\\
+\delta w -n\delta w+n  w^{5}-n\sigma w ^{3}+4n w^{2}w_{zz} +nw w_{zzz} -  w^{5}+
\sigma\,  w^{3}+\\+\sigma w_{zz} -5\,  w_{zzz} w -10\, w_{zz}   w^{2}-w_{zzzz}=0
\label{wBretherton}
\end{gathered}
\end{equation}

We suppose that solution of Eq.\eqref{wBretherton} can be presented in the form of Laurent series in an neighborhood of movable pole $z=z_{0}$
\begin{equation}
\begin{gathered}
w(z)=\sum\limits_{k=0}^{\infty}a_{k}\,(z-z_{0})^{k-p},\quad p>0
\label{L_expansion}
\end{gathered}
\end{equation}

The general solution of Eq.\eqref{wBretherton} has four different Laurent expansions in a neighborhood of the movable first order ($p=1$) pole $z=z_0$. Taking the autonomous of equation \eqref{wBretherton} into account we can take $z_0=0$. Substituting series \eqref{L_expansion} into Eq. \eqref{wBretherton} we have four following values of $a_{0}$
\begin{equation}
a_{0}^{(1)}=1,\quad a_{0}^{(2)}=2, \quad a_{0}^{(3)}=3, \quad a_{0}^{(4)}=-\frac{4}{n-1}.
\end{equation}

The necessary conditions for the existence of the elliptic solutions of equation Eq.\eqref{wBretherton} we can find from the following equations \cite{Kudryashov2010, Kudryashov2010a, Hone01, Eremenko01}
\begin{equation}
a_{0}^{(i)}+a_{0}^{(4)}=0, \qquad (i=1,2,3),
\label{E}
\end{equation}
\begin{equation}
a_{0}^{(i)}+a_{0}^{(j)}+a_{0}^{(4)}=0, \qquad (i,j=1,2,3),\, \qquad i\neq j,
\label{E1}
\end{equation}

\begin{equation}
a_{0}^{(1)}+a_{0}^{(2)}+a_{0}^{(3)}+a_{0}^{(4)}=0.
\label{E2}
\end{equation}

Solving equations \eqref{E}, \eqref{E1} and \eqref{E2} we obtain the following values  $n$ for existence
of the elliptic solutions of Eq.\eqref{wBretherton}
\begin{equation}
n=\frac{5}{3}, \quad n=\frac{9}{5}, \quad n=2,\quad n=\frac{7}{3}, \quad n=3,\quad n=5
\end{equation}


Cases $n=2$, $n=3$ and $n=5$ were investigated in \cite{Kudryashov2010a}. Let us consider the cases of $n=\frac{5}{3}$, $n=\frac{9}{5}$ and $n=\frac{7}{3}$.

Note that we should consider the Laurent series for  solution  of Eq. \eqref{wBretherton} corresponding to coefficient $a_{0}^{(4)}$ \cite{Kudryashov2009}.

Consider the case $n=\frac{5}{3}$. The corresponding values of $a_{0}^{(i)}$ are following
\begin{equation}
a_{0}^{(1)}=1,\quad a_{0}^{(2)}=2, \quad a_{0}^{(3)}=3, \quad a_{0}^{(4)}=-6
\end{equation}
We see that elliptic solution may be obtained if and only if we use the solution corresponding to combination of the four asymptotic expansions.

Following to the method \cite{Kudryashov2010,Kudryashov2010a} we construct the formal Laurent expansion of Eq.\eqref{wBretherton} solution corresponding to coefficient $a_{0}^{(4)}=-6$
\begin{equation}
\begin{gathered}
w=-\frac{6}{z}-\frac {\sigma}{50}z- \frac {50\delta+107\sigma^{2} }{615000} z^{3}-\vspace{0.1cm}\\
\\
-{\frac {\sigma \left( 209\,{\sigma}^{2}+900\,\delta \right)}{129150000}} z^{5}+\ldots+a_{16}\,z^{15}+\ldots
\label{53_expansion}
\end{gathered}
\end{equation}

Let us note that coefficient $a_{16}$ in the Laurent series \eqref{53_expansion} is arbitrary.
In accordance with the method \cite{Kudryashov2010,Kudryashov2010a} the general elliptic solutions has the following form
\begin{equation}
w=\frac{a_{0}^{(1)}(\wp_{z}+B_{1})}{2(\wp-A_{1})}+\frac{a_{0}^{(2)}(\wp_{z}+B_{2})}
{2(\wp-A_{2})}+\frac{a_{0}^{(3)}(\wp_{z}+B_{3})}{2(\wp-A_{3})}+h_{0}
\label{sW2}
\end{equation}
where $\wp(z,g_2,g_3)$ is the Weierstrass elliptic function.

Now we must compare expansion \eqref{53_expansion} with the Laurent expansion for expression \eqref{sW2}. The Laurent expansion for \eqref{sW2} may be obtained by using addition formulas for $\wp$ and $\wp_{z}$. The Laurent series and addition formulas for $\wp$ and $\wp_{z}$ are given in \cite{Abramowitz}. Also the Laurent expansion for expression \eqref{sW2} may be obtained with help of symbolic computation software Maple.

Comparing expansion \eqref{53_expansion} with the Laurent expansion for expressions \eqref{sW2} we get
\begin{equation}
\begin{gathered}
A_{1}=A_{2}=A_{3}=\frac{\sigma}{300},\quad B_{1}=-\frac{3B_{2}+B_{3}}{2},\quad \delta=-\frac{144\sigma^{2}}{625},\\
g_{2}=\frac{\sigma^{2}}{7500},\quad g_{3}=-\frac{\sigma^{3}}{3375000}, \,\, a_{16}=\frac{3617\sigma^{8}}{271367971875000000000000000}
\label{nC1}
\end{gathered}
\end{equation}

We see that there is constraint on the coefficient $a_{16}$. Otherwise solution in the form \eqref{sW2} does not exist.

However at these values of invariants $g_{2},g_{3}$ given in \eqref{nC1} the Weierstrass elliptic function is degenerated to trigonometric function
\begin{equation}
\begin{gathered}
\wp\left(z,\frac{\sigma^{2}}{7500},-\frac{\sigma^{3}}{3375000}\right)=-\frac{\sigma}
{150}\left(1-\frac{3}{2}\tanh^{2}\left\{\frac{\sqrt{\sigma}}{10}z\right\}\right)
\label{WpD_2}
\end{gathered}
\end{equation}

Substituting formulae \eqref{WpD_2} into \eqref{sW2} and taking \eqref{nC1} into account we have solution of the Eq. \eqref{wBretherton} in the from
\begin{equation}
\begin{gathered}
w=-\frac{3}{5}\sqrt{\sigma}\tanh\left\{\frac{\sqrt{\sigma}}{10}z\right\}
\end{gathered}
\end{equation}
We can see that solution \eqref{sW2} is degenerated to the simple periodic solution.

In this case the exact solution of the Eq.\eqref{rrBretherton_2} has the following form
\begin{equation}
\begin{gathered}
y=\frac{189^{3/2}}{15625}\,\sigma^{3}\,\cosh^{-6}\left(\frac{\sqrt{\sigma}}{10}\,z\right)
\end{gathered}
\end{equation}



In the case of $n=\frac{9}{5}$ values of $a_{0}^{(i)}$ are following
\begin{equation}
a_{0}^{(1)}=1,\quad a_{0}^{(2)}=2, \quad a_{0}^{(3)}=3, \quad a_{0}^{(4)}=-5
\end{equation}
Therefore elliptic solution may be obtained if and only if when we consider solution corresponding to combination of $a_{0}^{(2)},\,a_{0}^{(3)},\,a_{0}^{(4)}$.

Accordingly with method from \cite{Kudryashov2010,Kudryashov2010a} the solution can have the following form
\begin{equation}
w=\frac{a_{0}^{(2)}(\wp_{z}+B_{1})}{2(\wp-A_{1})}+\frac{a_{0}^{(3)}(\wp_{z}+B_{2})}{2(\wp-A_{2})}+h_{0}
\label{sW1}
\end{equation}

As in the previous case we construct the formal Laurent expansion of Eq.\eqref{wBretherton} solution  at $n=\frac{9}{5}$ corresponding to coefficient  $a_{0}^{(4)}=-5$
\begin{equation}
\begin{gathered}
w=-\frac{5}{z}-\frac {5\sigma}{222}z- \left( \frac {137{\sigma}^{2}}{492840}
+\frac {\delta}{750} \right) {z}^{3}-\\
\\
-
\frac {\sigma\, \left( 72775\,{\sigma}^{2}+323084\,\delta \right)}{19146834000} z^{5}+\ldots+a_{14}\,z^{13}+\ldots,
 \label{nL1}
\end{gathered}
\end{equation}
where coefficient $a_{14}$ is an arbitrary constant.

The Laurent expansion for \eqref{sW1} can be calculated using formulas from textbook \cite{Abramowitz} or using Maple command. Comparing the expansion \eqref{nL1} with the Laurent expansion for function \eqref{sW1} we have
\begin{equation}
\begin{gathered}
h_{0}=0,\quad A_{1}=A_{2}=\frac{\sigma}{222}, \quad B_{1}=-\frac{3B_{2}}{2}, \\
\delta=-\frac{1225\,\sigma^{2}}{5476},\quad g_{3}=-\frac{\sigma^{3}}{1367631}, \quad g_{2}=\frac{\sigma^{2}}{4107},\\ a_{14}=-\frac{\sigma^{7}}{11083927002941913120}.
\label{Tc2}
\end{gathered}
\end{equation}
We obtain constraint on the coefficient $a_{14}$. Only in this case Laurent series corresponding solution \eqref{sW1} exist.

However at the values of parameters \eqref{Tc2} elliptic solution \eqref{sW1} degenerate to the following simple periodic solution
\begin{equation}
\begin{gathered}
w=-5\sqrt{\frac{\sigma}{74}}\tanh \left( \sqrt{\frac{\sigma}{74}} \left( z-z_{0} \right) \right)
\end{gathered}
\end{equation}

In this case the exact solution of the Eq.\eqref{rrBretherton_2} has following form
\begin{equation}
\begin{gathered}
y=\frac{420^{5/4}\,1369^{3/4}}{1874161}\,\sigma^{5/2}\,\cosh^{-5}\left\{\sqrt{\frac{\sigma}{74}}\,z\right\}
\end{gathered}
\end{equation}

Let us consider the case of $n=\frac{7}{3}$. As this takes place $a_{0}^{(i)}$ have the following values
\begin{equation}
a_{0}^{(1)}=1,\quad a_{0}^{(2)}=2, \quad a_{0}^{(3)}=3, \quad a_{0}^{(4)}=-3
\end{equation}

We see that elliptic solution may be obtained if we consider solutions corresponding to combination of $a_{0}^{(3)}=3$ and $a_{0}^{(4)}$ or $a_{0}^{(1)},\,a_{0}^{(2)}$ and $ a_{0}^{(4)}$.

Consider the solution correspond to  coefficients $a_{0}^{(3)}=3$ and $a_{0}^{(4)}$. Accordingly to method from papers \cite{Kudryashov2010,Kudryashov2010a} we construct the formal Laurent expansion of solution for Eq.\eqref{wBretherton} at $n=\frac{7}{3}$  corresponding $a_{0}^{(4)}=-3$
\begin{equation}
\begin{gathered}
w=-\frac{3}{z}-\frac{\sigma}{34}z- \left( \frac {\delta}{210}+\frac
{211\sigma^{2}}{242760} \right) z^{3}-\vspace{0.1cm}\\
\\
-\frac {\sigma\, \left( 679\,{\sigma}^{2}+3468\,\delta \right)}{24761520} z^{5}+\ldots+a_{10}\,z^{9}+\ldots
 \label{nL}
\end{gathered}
\end{equation}

Note that coefficient $a_{10}$ in series \eqref{nL} is arbitrary constant.

Following \cite{Kudryashov2010,Kudryashov2010a} we look for the elliptic solution of Eq.\eqref{wBretherton} in the form
\begin{equation}
w=\frac{a_{0}^{(3)}(\wp_{z}+B)}{2(\wp-A)}+h_{0}
\label{sW_n}
\end{equation}

Comparing the expansion \eqref{nL} with the Laurent expansion for expression \eqref{sW_n} we have
\begin{equation}
\begin{gathered}
h_{0}=0,\quad A=\frac{\sigma}{102}, \quad B=0, \quad \delta=-\frac{225\,\sigma^{2}}{1156},\\
a_{10}=-\frac{\sigma^{5}}{708451848720},\quad g_{3}=-\frac{\sigma^{3}}{132651}, \quad g_{2}=\frac{\sigma^{2}}{867},\\
 a_{10}=-\frac{\sigma^{5}}{708451848720}.
\end{gathered}
\end{equation}

We see that there is constraint on the coefficient $a_{10}$. Otherwise solution in the form \eqref{sW2} does not exist.

In this case the Weierstrass elliptic function is degenerated to trigonometric function as well
\begin{equation}
\begin{gathered}
\wp(z-z_{0},\frac{\sigma^{2}}{867},-\frac{\sigma^{3}}{132651})=\frac{\sigma}{102}\left(3\,
\tan^{2}\left\{\sqrt{\frac{\sigma}{34}}(z-z_{0})\right\}-2\right)
\label{WpD}
\end{gathered}
\end{equation}

After substituting formulae \eqref{WpD} into \eqref{sW_n} we have solution of \eqref{wBretherton} in the form
\begin{equation}
w=-3\,\sqrt{\frac{\sigma}{34}}\,\tanh\left\{\sqrt{\frac{\sigma}{34}}(z-z_{0})\right\}.
\label{Solution}
\end{equation}

Using the B\"{a}cklund transformations \eqref{Bck} we obtain exact solution of the Eq.\eqref{rrBretherton_2} in the form
\begin{equation}
\begin{gathered}
y=\left(\frac{90}{289}\right)^{3/4}\,\sigma^{3/2}\,\cosh^{-3}\left\{\sqrt{\frac{\sigma}{34}}\,z\right\}
\end{gathered}
\end{equation}

In the case of $n=7/3$ and solution corresponding to combination to three asymptotic expansions $a_{0}^{(1)},a_{0}^{(2)}, a_{0}^{(4)}$ exactly the same solution as \eqref{Solution}.

So we see that  Eq. \eqref{rrBretherton_2} has the elliptic solutions only at $n=2$, $n=3$  and $n=5$. This completes the prof. $\Box$

{\bf{Remark 1}.} Let us consider the case of $\sigma=0$ separately.
Comparing the expansion \eqref{nL} at $\sigma=0$ with the Laurent expansion for function \eqref{sW_n} in the case of $n=\frac{7}{3}$ we have
\begin{equation}
\begin{gathered}
A=B=h_{0}=0,\quad a_{10}=-\frac{\delta^{3}}{2167074000}, \quad g_{2}=-\frac{\delta}{63}, \quad g_{3}=0
\end{gathered}
\end{equation}

Using this relation we obtain the elliptic solution of Eq.\eqref{wBretherton} in the form
\begin{equation}
\begin{gathered}
w=- \frac{3\,\wp_{z}\left( z,-{\frac {\delta}{63}},0
 \right) }{2\,\wp \left( z,-{\frac {\delta}{63}},0 \right) }
 \label{R1_S}
\end{gathered}
\end{equation}
The case of $\sigma=0$ means that $C_{0}^{2}=\alpha$. So we can see that elliptic solutions exist only at  special value of $C_{0}$.

Using the B\"{a}cklund transformations \eqref{Bck} from  \eqref{R1_S} we obtain the elliptic solution of Eq.\eqref{rrBretherton_2} in the form
\begin{equation}
\begin{gathered}
y=\sqrt[4]{360}\,\left[\wp(z,-\frac{\delta}{63},0)\right]^{3/2}
\end{gathered}
\end{equation}

In the cases $n=\frac{5}{3}$ and $n=\frac{9}{5}$ at $\sigma=0$ elliptic solutions are degenerated to the rational solutions under condition $\delta=0$.

{\bf{Remark 2}} The simple periodic solution of Eq. \eqref{wBretherton} has the following form
\begin{equation}
w=-\frac{4\pi}{(n-1)\,T}\,\cot\left(\frac{\pi z}{T}\right)
\label{nPeriodic}
\end{equation}
where $T$ and $\sigma$ have values
\begin{equation}
\begin{gathered}
T_{1,2}=\pm\frac{2\,\sqrt{-2\sqrt{-\delta}(n+1)}\pi}{(n-1)\sqrt{-\delta}}, \quad
T_{3,4}=\pm \frac{2\,\sqrt{2\sqrt{-\delta}(n+1)}\pi}{(n-1)\sqrt{-\delta}},\\
\sigma_{1,2}=-\frac{\delta(n^{2}+2n+5)}{2\sqrt{-\delta}(n+1)},\quad \sigma_{3,4}=\frac{\delta(n^{2}+2n+5)}{2\sqrt{-\delta}(n+1)}
\label{R2_C}
\end{gathered}
\end{equation}
Using formulae \eqref{nPeriodic}, relations \eqref{R2_C} and taking B\"{a}cklund transformations \eqref{Bck} into account we get
\begin{equation}
\begin{gathered}
y_{1,2}=\left(-\frac{(3n^{2}+10n+3)\delta}{8(n+1)\left[\cos\left\{\frac{(n-1)\sqrt{-2\delta}\,z}
{4\sqrt{\mp(n+1)\sqrt{-\delta}}}\right\}-1\right]^{2}}\right)^{\frac{1}{n-1}}
\label{nSP}
\end{gathered}
\end{equation}
Expression \eqref{nSP} is the simple periodic solution of Eq. \eqref{rrBretherton_2} at any $n\neq \pm 1$.

\section{Exact solutions of Eq. \eqref{rBretherton} in the case of $n=5$ and $m\neq1$.}

Suppose that solutions of Eq. \eqref{rBretherton} have the asymptotic expansions in the form of the Laurent series in a neighborhood of the pole $z=z_{0}$
\begin{equation}
y(z)=\sum\limits_{k=0}^{\infty}\,a_{k}\,(z-z_{0})^{k-p}, \quad p>0
\label{L3}
\end{equation}

Eq.\eqref{rBretherton} is autonomous and without loss of the generality we can assume $z_0=0$.

In the case of  $n=5$ Eq. \eqref{rBretherton} has the four different formal Laurent expansions in a neighborhood pole $z=0$ of the first order ($p=1$).

Without loss of generality we use the following values of parameters $\gamma$ and $\beta$
\begin{equation}
\beta=-1, \quad \gamma=24
\end{equation}

Corresponding values of the coefficients $a_{0}$ are the following
\begin{equation}
a_{0}^{(1)}=1,\quad a_{0}^{(2)}=-1, \quad a_{0}^{(3)}=i, \quad a_{0}^{(4)}=-i.
\end{equation}

The necessary conditions for existence of the elliptic solutions take the form
\begin{equation}
a_{0}^{(1)}+a_{0}^{(2)}=0 \quad \mbox{and} \quad a_{0}^{(3)}+a_{0}^{(4)}=0
\label{E3}
\end{equation}

We see that the coefficients $a_{0}^{(1)}$ and $a_{0}^{(3)}$ correspond to the elliptic solutions.

The Fuchs indices corresponding of each expansions are following
\begin{equation}
j_{1}=-1,\quad j_{2,3}=\frac{1}{2}\left(5\pm\sqrt{39}\right)
\end{equation}
So we see that coefficients of expansion \eqref{L3} for the solution of  Eq. \eqref{rBretherton} are unique.

We use the method from the references \cite{Kudryashov2010,Kudryashov2010a} for constructing the elliptic solutions Eq. \eqref{rBretherton}. The algorithm of this method is the following.

1) Construct  the formal Laurent expansion of solution for Eq.\eqref{rBretherton};

2) Take the general form of the possible elliptic for Eq.\eqref{rBretherton} and find the formal Laurent expansion for this form of solution;

3) Compare the formal Laurent expansion of solution for Eq.\eqref{rBretherton} hat was found on the first step with the formal Laurent expansion of the possible elliptic solution that was found on the second step;

4) Solve the system of the algebraic equations obtained on the third step and find the parameters of the  Eq.\eqref{rBretherton_1} and parameter of the possible elliptic solution.

{\bf Remark 1.} The formal Laurent expansion for general form of the possible elliptic solution can be found by using textbook \cite{Abramowitz} or by using Maple.

{\bf Remark 2.} The simple periodic solution can be found by using analogues algorithm. See details in references \cite{Kudryashov2010,Kudryashov2010a}.

Consider the exact solutions of Eq. \eqref{rBretherton} in the case $m=2$.

Substituting $m=2$ into Eq. \eqref{rBretherton} in the case of $n=5$ we have
\begin{equation}
\frac{\alpha-C_{0}^{2}}{2}\,y_{z}^{2}-(y_{z}\,y_{zzz}-\frac{1}{2}\,y_{zz}^{2})+
\frac{\delta}{3}\,y^{3}+4\,y^{6}+C_{2}=0
\label{rBretherton_1}
\end{equation}

The formal Laurent expansion of the solution for Eq.\eqref{rBretherton_1} corresponding to $a_{0}^{(1)}=1$  is the following
\begin{equation}
\begin{gathered}
y=\frac{1}{z}+ \left(\frac{\alpha-C_{0}^{2}}{60}\right) z-\frac {\delta}{120}\,{z}^{2}-\frac {\left( C_{0}^{2}-\alpha \right) ^{2}}{1800}z^{3}-\frac {\delta\, \left( C_{0}^{2}-\alpha \right)}{5760} z^{4}+\ldots
 \label{L}
\end{gathered}
\end{equation}

The possible exact solution of Eq. \eqref{rBretherton_1} takes the form
\begin{equation}
y=\frac{a_{0}^{(2)}(\wp_{z}+B)}{2(\wp-A)}+h_{0}
\label{sW}
\end{equation}
where $\wp$ is Weierstrass elliptic function.

Comparing the expansion \eqref{L} with the Laurent expansion of expression \eqref{sW} we have
\begin{equation}
\begin{gathered}
h_{0}=0,\quad A=0, \quad B=\frac{\delta}{60}, \quad C_{0}=\pm\sqrt{\alpha},\\
g_{2}=0,\quad g_{3}=-\frac{\delta^{2}}{3600}, \quad C_{2}=-\frac{\delta^{2}}{1800}
\end{gathered}
\end{equation}

Using these relations we obtain the elliptic solution of Eq.\eqref{rBretherton_1} in the form
\begin{equation}
\begin{gathered}
y=-\frac{\wp'(z,0,-\frac{\delta^{2}}{3600})+\frac{\delta}{60}}{2\wp(z,0,-
\frac{\delta^{2}}{3600})}
\end{gathered}
\end{equation}

Let us look for the exact solutions of Eq. \eqref{rBretherton} in the case of $m=3$.
Substituting $m=3$ into Eq. \eqref{rBretherton} in the case of $n=5$ we have
\begin{equation}
\frac{\alpha-C_{0}^{2}}{2}\,y_{z}^{2}-(y_{z}\,y_{zzz}-\frac{1}{2}\,y_{zz}^{2})+
\frac{\delta}{4}\,y^{4}+4\,y^{6}+C_{2}=0
\label{rBretherton_2}
\end{equation}

The formal Laurent expansion of the solution for Eq.\eqref{rBretherton_1} corresponding to $a_{0}^{(1)}=1$ is the following
\begin{equation}
\begin{gathered}
y=\frac{1}{z}+ \left(\frac{2(\alpha-C_{0}^{2})-\delta}{120} \right) z-
\frac {\left(2(C_{0}^{2}-\alpha)+\delta \right)  \left(4(
C_{0}^{2}-\alpha)-\delta \right) }{14400} z^{3}+\ldots
 \label{L1}
\end{gathered}
\end{equation}
Comparing the expansion \eqref{L1} with the Laurent expansion for the expression \eqref{sW} we have
\begin{equation}
\begin{gathered}
A=\frac{2(C_{0}^{2}-\alpha)-\delta}{120},\quad B=h_{0}=0, \quad g_{2}=\frac{(C_{0}^{2}-\alpha)(2(C_{0}^{2}-\alpha)-\delta)}{240},\vspace{0.2cm} \\
g_{3}=\frac {\left( 13\, C_{0}^{2}+\delta-13\,\alpha \right)  \left( -2\,\alpha+2\, C_{0}^{2}-\delta \right) ^{2}}{432000},\vspace{0.2cm} \\
C_{2}=-\frac {\left(2(C_{0}^{2}-\alpha)-\delta \right)
\left(8(C_{0}^{2}-\alpha) +\delta\right)  \left(18(C_{0}^{2}-\alpha)+\delta \right)}{96000}.
\label{m=3_coeffs}
\end{gathered}
\end{equation}
Using these relations we obtain the elliptic solutions of Eq.\eqref{rBretherton_2} in the form
\begin{equation}
\begin{gathered}
y=-\frac{\wp'(z-z_{0},g_{2},g_{3})}{2\wp(z-z_{0},g_{2},g_{3})-\frac{2(\alpha-C_{0}^{2})-\delta}{60}}
\end{gathered}
\end{equation}
where $g_{2},g_{3}$ are determined by the relations \eqref{m=3_coeffs}

Let us look for the periodic solutions of Eq.\eqref{rBretherton_2}. In accordance with the method from the references \cite{Kudryashov2010,Kudryashov2010a} we search for the periodic solutions in the form
\begin{equation}
\begin{gathered}
y=-\sqrt{L}\frac{A\cot(\sqrt{L}z)+\sqrt{L}}{A-\sqrt{L}\cot(\sqrt{L}z)}+\sqrt{L}\cot(\sqrt{L}z)+h_{0}
\label{P}
\end{gathered}
\end{equation}
Comparing the expansion \eqref{L1} with the Laurent expansion for the possible solution \eqref{P} we have
\begin{equation}
\begin{gathered}
\delta=3(\alpha-C_{0}^{2}),\quad L=\frac{\alpha-C_{0}^{2}}{8}, \quad A=\sqrt{-L}, \\ h_{0}=-\sqrt{-L},\quad C_{2}=-L^{3}
\end{gathered}
\end{equation}

Let us consider the exact solutions of Eq. \eqref{rBretherton} in the case of $m=4$.
Substituting $m=4$ into Eq. \eqref{rBretherton} in the case $n=5$ we have
\begin{equation}
\frac{\alpha-C_{0}^{2}}{2}\,y_{z}^{2}-(y_{z}\,y_{zzz}-\frac{1}{2}\,y_{zz}^{2})+
\frac{\delta}{5}\,y^{5}+4\,y^{6}+C_{2}=0
\label{rBretherton_3}
\end{equation}
The formal Laurent expansion of solution for Eq.\eqref{rBretherton_1} corresponding to coefficient $a_{0}^{(1)}=1$ is the following
\begin{equation}
\begin{gathered}
y=\frac{1}{z}-\frac {\delta}{120}+ \left( \frac{120(\alpha-C_{0}^{2})+\delta^{2}}{7200}\right) z
-\frac {\delta^{3}}{432000}{z}^{2}+\ldots
 \label{L2}
\end{gathered}
\end{equation}
Comparing the expansion \eqref{L1} with the Laurent expansion for the expression \eqref{sW} we have
\begin{equation}
\begin{gathered}
A=\frac{7\delta^{2}}{43200},\quad B=\frac{\delta^{3}}{216000}, \quad h_{0}=-\frac{\delta}{120},\vspace{0.2cm}  \\ g_{2}=-\frac {{\delta}^{4}}{31104000},\quad
g_{3}=\frac {\delta^{6}}{1259712000000},\vspace{0.2cm} \\
C_{2}=-\frac {\delta^{6}}{186624000000}, \quad C_{0}=\pm\frac{1}{12}\sqrt{\frac{720\alpha-\delta^{2}}{5}}
\label{m=4_coeffs}
\end{gathered}
\end{equation}
Using these relations we obtain the elliptic solutions of Eq.\eqref{rBretherton_1} in the form
\begin{equation}
\begin{gathered}
y=-\frac{\wp'(z-z_{0},g_{2},g_{3})+\frac{\delta^{3}}{216000}}{2\wp(z-z_{0},g_{2},g_{3})-
\frac{7\delta^{2}}{21600}}-\frac{\delta}{120}
\end{gathered}
\end{equation}
where $g_{2},g_{3}$ are determined by relations \eqref{m=4_coeffs}.

\section{Exact solutions of Eq. \eqref{rBretherton} at $m=2$ in the case $n=3$.}

Let us consider the exact solutions of Eq. \eqref{rBretherton} in the case of $n=3$ and $m=2$.
The case of $n=3$ and $m=1$ was considered in \cite{Kudryashov2010a}. From Eq. \eqref{rBretherton} we have
\begin{equation}
\frac{\alpha-C_{0}^{2}}{2}\,y_{z}^{2}+\beta(y_{z}\,y_{zzz}-\frac{1}{2}\,y_{zz}^{2})+
\frac{\delta}{3}\,y^{3}+\frac{\gamma}{4}\,y^{4}+C_{2}=0.
\label{rBretherton_5}
\end{equation}

Without loss of the generality we take the following values of $\beta$ and $\gamma$
\begin{equation}
\beta=-1, \quad  \gamma=120.
\end{equation}

We use the algorithm  again from the reverences \cite{Kudryashov2010,Kudryashov2010a} that we have  described in the previous section.

We assume that the solution of  Eq.\eqref{rBretherton_5} can be presented in the form of the Laurent series \eqref{L3}. Without loss of the generality we set $z_{0}=0$.

Eq. \eqref{rBretherton_5} admits two different Laurent expansions in the neighborhood of the pole $z=0$
\begin{equation}
y=\frac{1}{z^{2}}+\frac {6(C_{0}^{2}-\alpha)-\delta}{360}-\frac {\left( -\delta+6\,C_{0}^{2}-6\,\alpha \right)  \left( \delta+6\,C_{0}^{2}-6\,\alpha
 \right) }{129600} z^{2}+\ldots
\label{n=3_L1}
\end{equation}
\begin{equation}
y=-\frac{1}{z^{2}}+\frac {6(\alpha-C_{0}^{2})-\delta}{360}+\frac{ \left( -\delta+6\,C_{0}^{2}-6\,\alpha \right)  \left( \delta+6\,C_{0}^{2}-6\,\alpha
 \right)}{129600} z^{2}+\ldots
\label{n=3_L2}
\end{equation}

We see that $a_{-1}^{(1)}$ and $a_{-1}^{(2)}$ are zeros. So the necessary condition for the existence of the elliptic solution is hold.

We use the expression for the solution of Eq.\eqref{rBretherton_5} in the following form
\begin{equation}
y=2\wp(z,g_{2},g_{3})-\frac{1}{4}\left[\frac{\wp'(z,g_{2},g_{3})+B}{\wp(z,g_{2},g_{3})-A}\right]^{2}+h_{0}+A
\label{n=3_Wp}
\end{equation}

Comparing the expansion \eqref{n=3_L1}  with the Laurent expansions to \eqref{n=3_Wp}  we have
\begin{equation}
\begin{gathered}
A=\frac{\alpha-C_{0}^{2}}{60},\quad B=0, \quad g_{2}=\frac{(C_{0}^{2}-\alpha)^{2}}{540}, \quad g_{3}=\frac{(C_{0}^{2}-\alpha)^{3}}{81000},\\
\\
C_{2}=-\frac{11(C_{0}^{2}-\alpha)^{4}}{303750}.
\end{gathered}
\end{equation}

After substituting this relations in formulae \eqref{n=3_Wp} we get following solution of Eq.\eqref{rBretherton_5}
\begin{equation}
\begin{gathered}
y=2\,\wp \left( z,\frac { \left( C_{0}^{2}-\alpha \right) ^{2}}{540},\frac {\left( C_{0}^{2}-\alpha \right) ^{3} }{81000}\right)-\vspace{0.1cm} \\
-\frac{1}{4}\, \left[\frac{  \wp' \left( z,\frac {\left( C_{0}^{2}-\alpha \right) ^{2}}{540},\frac {\left( C_{0}^{2}-\alpha \right)^{3}}{81000} \right)  }{  \wp \left( z,\frac { \left( C_{0}^{2}-\alpha \right) ^{2}}{540},\frac {\left( C_{0}^{2}-\alpha \right) ^{3} }{81000}\right) +\frac{C_{0}^{2}-\alpha}{60} } \right]^{2}+\frac {\alpha-C_{0}^{2}}{60}
\end{gathered}
\end{equation}

We should note that this solution can be obtained from solution (4.19) in the reference \cite{Kudryashov2010a} assuming $C_{0}=0$.

\section{Conclusion}

In this paper we have studied all possible solutions of the generalized Bretherton equation. The meromorphic exact solutions of this equation are classified in the cases $n=5$ at $m=2$, $m=3$  and $m=4$. The case $m=1$ and arbitrary $n$ also were considered. We have shown that the elliptic solutions of the generalized Bretherton equation exist only in the cases $n=2$, $n=3$ and $n=5$ at $\sigma\neq0$ and at $\delta\neq0$. We have obtained that there is the elliptic solution in the case $n=\frac73$ at  $C_{0}^{2}=\alpha$.

\end{document}